\documentclass[a4paper,fleqn,usenatbib]{mnras}

\usepackage[T1]{fontenc}
\usepackage{ae,aecompl}

\usepackage{graphicx}	
\usepackage{amsmath}	
\usepackage{amssymb}	

\title[Volcanic activity and the exosphere of HD3167b]
{Searching for volcanic activity and a Mercury-like exosphere of the
  super-Earth HD3167 b}

\author[Eike W. Guenther \& Kristina G. Kislyakova]{ Eike
  W. Guenther,$^{1}$\thanks{E-mail: guenther@tls-tautenburg.de}
  Kristina G. Kislyakova$^{2}$ \\
$^{1}$ Th\"uringer Landessternwarte Tautenburg, Sternwarte 5, 07778
  Tautenburg, Germany\\
$^{2}$ Department of Astrophysics, University of Vienna,
T\"urkenschanzstrasse 17, 1180 Vienna, Austria\\
}

\date{Accepted XXX. Received 2019 October 16; in original form 2019 August 29}

\pubyear{2019}

\begin{document}
\label{firstpage}
\pagerange{\pageref{firstpage}--\pageref{lastpage}}
\maketitle


\begin{abstract}
HD3167\,b is a transiting super-Earth that has a density which is
consistent with a rocky composition. The planet is exposed to strong
radiation, intense stellar wind, and likely strong tidal forces
and induction heating. According to theory, planets that are so
close to the star should have an atmosphere like Mercury but much more
extended and denser. Other theories predict that such planets have a
lava lake on their surfaces and exhibit an enormous volcanic
activity. We have calculated the heating by electromagnetic induction
to estimate if it can drive significant volcanic activity at HD3167\,b
and shown that for some magnetic fields the heating can be
substantial.  HD3167 is an ideal target to search for the exosphere of a
planet, and signs of volcanic activity.  We observed the planet in-
and out-of transit with UVES in order to search for presence of lines
originating from the exosphere of the planet such as the $\rm
Na\,D_{1,2}$ and Ca{\small II}\,H\&K lines as well as numerous [S II],
[S III], and [O III] lines that are tracers of volcanic activity.  We
derived upper limits for the ratios of the line fluxes to the stellar
flux. The upper limits that we derived are
$I_{p,\lambda}/I_{*,\lambda}=1.5\,10^{-3}$ for the Ca{\small II}\,H\&K
lines, and $I_{p,\lambda}/I_{*,\lambda}=7.2\,10^{-4}$ and
$I_{p,\lambda}/I_{*,\lambda}=3.3\,10^{-4}$ for the $\rm NaD_{1,2}$
lines, respectively. The fact that our upper limits correspond to
previous detections in 55\,Cnc\,e shows that not all super-Earth show
these lines all the time and that they might be variable.
\end{abstract}




\section{Introduction}
\label{sectI}

After the detection of giant exoplanets, the most exciting discovery
was the detection of planets with masses of less than 15 $\rm
M_{Earth}$ that have densities consistent with a rocky composition.  A
special class of such planets are transiting ultra-short-period
planets which have orbital periods of one day or less.  These planets
have the advantage that due to their short periods, their masses and
radii can be determined with high accuracy.  Currently known
transiting ultra-short-period planets (USPs) include CoRoT-7b
\citep{leger09}, 55\,Cnc\,e \citep{winn11}, Kepler-10b
\citep{batalha11}, Kepler-78b \citep{sanchis13}, WASP-47e
\citep{dai15,sinukoff17a}, KOI 1843.03 \citep{rappaport13}, K2-106
\citep{guenther17}, and HD3167b (discovery: \cite{vanderburg16},
  characterization: \cite{gandolfi17}).

Although these planets are often called "super-Earths", the question
is how Earth-like are they really? USPs are certainly exposed to 
very strong radiation and stellar wind. Because volatiles of
  their atmospheres are efficiently removed, such planets 
  likely have only a thin atmosphere consisting of rocky vapors
  \citep{leger11,briot10,barnes10}. \citet{leger11}
  furthermore showed that the radiative as well as the tidal heating
  of these planets is so intense that it can drive extreme
  volcanism. Also, the temperature at the substellar point is so high
  that the surface rocks can melt and form a lava lake \citep{leger09,
    schaefer09, leger11}.  The lava lake model can also explain the
relatively high albedo of Kepler-10b (geometric albedo:
  $0.63\pm0.09$, Bond albedo: $0.91^{+0.06}_{-0.10}$;
  \citealp{hu15}), because a lava composed of $\rm ThO_2$ particles
dispersed in the $\rm Al_2\,O_3$ $\rm CaO$ would be almost as white as
snow in the visible light \citep{rouan11}.

This type of lava is more plausible than the lava we have on Earth,
because of its much higher melting point. Magma oceans are not only an
exotic feature of USPs since most, if not all, rocky planets had magma
ocean at a young age. Studying volcanic activity and lava oceans of
USPs is thus not only important for these types of planets but also
for understanding the evolution of rocky planets in general. For
example, according to theory the outgassing from these lava oceans
during their solidification produces $\rm CO_2$- and $\rm H_2 O$-rich
atmospheres \citep{elkins08,elkins12,lammer18}. According to
\cite{ito15} a planet with a rocky molten surface should have an
atmosphere that is composed of gas-phase species dominated by Na, K, Fe,
Si, SiO, O, and $\rm O_2$.

USPs are not only heated by tides and the radiation of the star. Their
interiors are also heated by electromagnetic induction if they are
embedded in varying in time magnetic fields of their host stars
\citep{kislyakova18,kislyakova17}. In some cases, induction heating
may even exceed the tidal heating. Since all stars with outer convective 
envelopes have magnetic fields, induction heating might be quite
important for many of the short-period planets, which  
may then lead to enhanced volcanic activity. 
However, the amount of outgasing also depends on the mass
of the planet. The highest level is expected for planets with masses
between 2--3~M$_\oplus$ \citep{dorn18}.

How can we possibly detect volcanic activity?  Possible tracers for
volcanism are the lines observed in Io and its plasma torus such as
the [S\,III] 3722, [O\,II] 3726, [O\,II] 3729, [S\,II] 4069, [S\,II]
4076, [O\,III] 5007, [O\,I] 6300, [S\,III] 6312, [S\,II] 6716, [S\,II]
6731 \AA ~lines, as well as the sodium D-lines \citep{brown74,
  brown75, brown76, brown82, morgan82, thomas93, thomas96, kueppers95,
  kueppers97, kupo76}. 

Some USPs are likely to have a Mercury-like exosphere, because light
volatiles like H, He, O, $\rm CH_4$ are quickly lost through the Jeans
and other escape mechanisms. Since these planets are much larger and
much closer to their host stars than Mercury, the amount of material
released will be enormous. How can we detect the exosphere? $\rm Ca^+$
has been detected in the Ca{\small II}\,H\&K lines in the exosphere
and coma of Mercury \citep{bida00} and the Na coma has been detected
during the Mercury transit in 2003 \citep{schleicher04}. Using this
scaled-up model of Mercury and applying it to the close-in super-Earth
CoRoT-7b, \cite{mura09} showed that CoRoT-7b should have a tail of
escaping $\rm Ca^+$ particles. With a length of 30 $\rm R_{planet}$,
this tail is expected to have a column density larger than $\rm
10^{16} m^{-2}$.  \cite{guenther17} observed CoRoT-7b in- and out-of
transit and obtained upper limits of the flux in Ca\,I, Ca\,II, and 
Na-lines that correspond to the most optimistic values in
\cite{mura09}. The main reason why neither volcanism nor the exosphere
had been detected in CoRoT-7b is likely due to a large distance to its host star
of $160.6\pm0.9$\,pc, (\citealp{chiappetti18}).

Luckily, a number of USPs have been detected that orbit much closer
stars. One of them is  55\,Cnc which has a distance of $12.590\pm0.012$\,pc
(\citealp{chiappetti18}). The orbital period of the USP 55\,Cnc\,e is is 0.74
days. Its mass is $8.37\pm0.38$ $\rm M_{Earth}$, and its radius 
$2.17\pm0.10$ $\rm R_{Earth}$. The density of 5\,Cnc\,e 
is $4.5\pm0.20$ $\rm g\,cm^{-3}$, which is roughly consistent with 
an Earth-like composition.
Using optical transmission spectroscopy, \cite{ridden16} searched for
an exosphere of this planet in the Na and singly ionized calcium $\rm
Ca^{+}$ lines in the high resolution spectra obtained with HARPS.
Combining the data of all five observed transits they detected a
signal potentially associated with sodium in the planet's exosphere at a
statistical significance level of 3$\sigma$.  Using the Ca{\small
  II}\,H\&K lines they also found a potential signal from ionized
calcium at the 4.1$\sigma$-level. Interestingly, this latter signal
originates from just one of the transit observations. This would
correspond to an optically thick $\rm Ca^+$ exosphere with the size of
approximately five Roche radii. If this were a real detection, it
would imply that the exosphere exhibits extreme variability. Besides
that, \cite{demory16} have used the middle IR wavelengths to search
for volcanic activity on 55\,Cnc\,e during secondary transits, with
very promising results.

Other potentially interesting objects to search for the signatures of
volcanism and an exosphere using optical transmission spectroscopy are
K2-106 \citep{guenther17} and HD3167\,b
\citep{vanderburg16,gandolfi17,christiansen17}.  What makes these
objects interesting is that their densities are higher than that of
55\,Cnc\,e.  K2-106\,b and HD3167\,b have a density of
$13.1^{-3.6}_{+5.4}$ and $8.00^{+1.10}_{-0.98}$ $\rm \,g~cm^{-3}$,
respectively. The density of HD3167\,b is consistent with a
composition containing up to 40\% iron and 60\% silicate, and
K2-106\,b to the composition containing possibly up to 70\% iron
\citep{gandolfi17,guenther17}.  For comparison, the Earth has an iron
contents of 32.5\% and Mercury of 70\%. HD3167\,b is the preferred
target, because its host star is 13 times brighter than CoRoT-7.
HD3167\,b has an orbital period of 0.96 days, a semi-major axis of
$0.01752\pm0.00063$\,AU, a mass of $5.69\pm0.44$ $\rm M_{Earth}$, a
radius of $1.574\pm0.054$ $\rm R_{Earth}$, and an equivalent
temperature of $\rm T_{eq}=1759\pm20\,K$.  The outer component,
HD3167\,c, has an orbital period of 29.8 days, a mass of
$8.33^{+1.79}_{-1.85}$ $\rm M_{Earth}$ and a radius of
$2.74^{+0.11}_{-1.00}$ $\rm R_{Earth}$ \citep{gandolfi17}.  The orbit
of HD3167\,c is highly inclined with respect to the stellar rotation
axis \citep{dalal19}. Because the orbital inclinations of the two
planets are rather similar with $\rm i=88.6^{+1.0}_{-1.4}\,deg$, and
$\rm i=89.6\pm0.2\,deg$, and since most systems are co-planar
\citep{winn15}, we assume that also HD3167\,b has a highly inclined
orbit. As we will discuss in this article, the highly inclined orbit
makes this planet particularly interesting for studies on the
planet-star interactions and induction heating.

According to \cite{christiansen17}, possibly a third non-transiting
planet, HD3167\,d, exists between the two transiting ones. The planet
is supposed to have an orbital period of 8.51 days and a K-amplitude
of $2.39\pm0.24$ $\rm ms^{-1}$, which corresponds to a minimum mass of
$6.90\pm0.71 $ $\rm R_{Earth}$. However, the radial-velocity (RV)
signal of this planet was not detected by \cite{gandolfi17}. Instead
they found two adjacent signals with periods of 6.0 and 10.8 days and
K-amplitudes of $1.34^{+0.27}_{-0.28}$ and $5.967^{+0.038}_{-0.035}$
$\rm ms^{-1}$. Whether there is a planet with a different inclination
between the two transiting ones is thus uncertain. The host is a
K0V-star with a rotation period of $23.52\pm2.87$\,d that has a
distance of only $47.4\pm0.2$\,pc \citep{chiappetti18}.

In this article we present the analysis of in-transit and
out-of-transit spectra of HD3167\,b obtained in order to search for
the exosphere and the volcanic activity of this planet. We discuss
observations in Section~\ref{sectII}, the data reduction in
Section~\ref{sectIIa}, and the results in Section~\ref{sectIII}. 
We discuss possible origins of
the atmosphere and calculate possible induction heating rates in
Section~\ref{sectIV}. Sections~\ref{sectV} contains the
discussion and Section~\ref{sectVI} our conclusions.

\section{Observations}
\label{sectII}

We observed HD3167 continuously on November 20, 2017 from UTC 00:34
until UTC 04:33 with UVES (Ultraviolet and Visual \'Echelle
Spectrograph) at the VLT (ESO program 099.C-0175(A)). This corresponds
to Heliocentric Julian Dates (HJD) from 2458077.52361 to
2458077.68958.  The transit occured from HJD 2458077.595261 to
2458077.664011 \citep{vanderburg16}.  UVES is the \'Echelle
spectrograph for the UT2 KUEYEN Telescope at the ESO Paranal
\citep{dekker00} and has two arms. The wavelength ranges of the two
arms are 3259 to 4493 \AA~(the "blue" arm) and 4726 to 6835 \AA~(the
"red" arm). The resolution of the spectra is $\lambda/\Delta \lambda =
52000$, or 5.5 $\rm kms^{-1}$, and the dispersion typically 1.0 $\rm
kms^{-1}pixels^{-1}$.  During the four hours of observations 58
spectra were obtained in the blue channel and 65 in the red
channel. The exposure times in the blue and red channel were 200~s, and
200~s (first 13 franes) and 150~s (all other frames), respectively.

\section{Reduction of the data}
\label{sectIIa}

The standard EsoReflex pipeline was used for the basic data reduction
which includes the bias subtraction, the flat-fielding, the extraction
of the spectra and the wavelength-calibration \citep{moeller19}.
Additionally to these, we also reduced the spectra using IRAF (Image
Reduction and Analysis Facility) to find out whether that makes any
difference. Both types of reduced spectra were essentially identical.
After these initial steps have been done, we carried out seven
additional steps to extract the signal of the planet. We describe this
procedure below.

\begin{figure}
\includegraphics[height=0.25\textheight,angle=0.0]{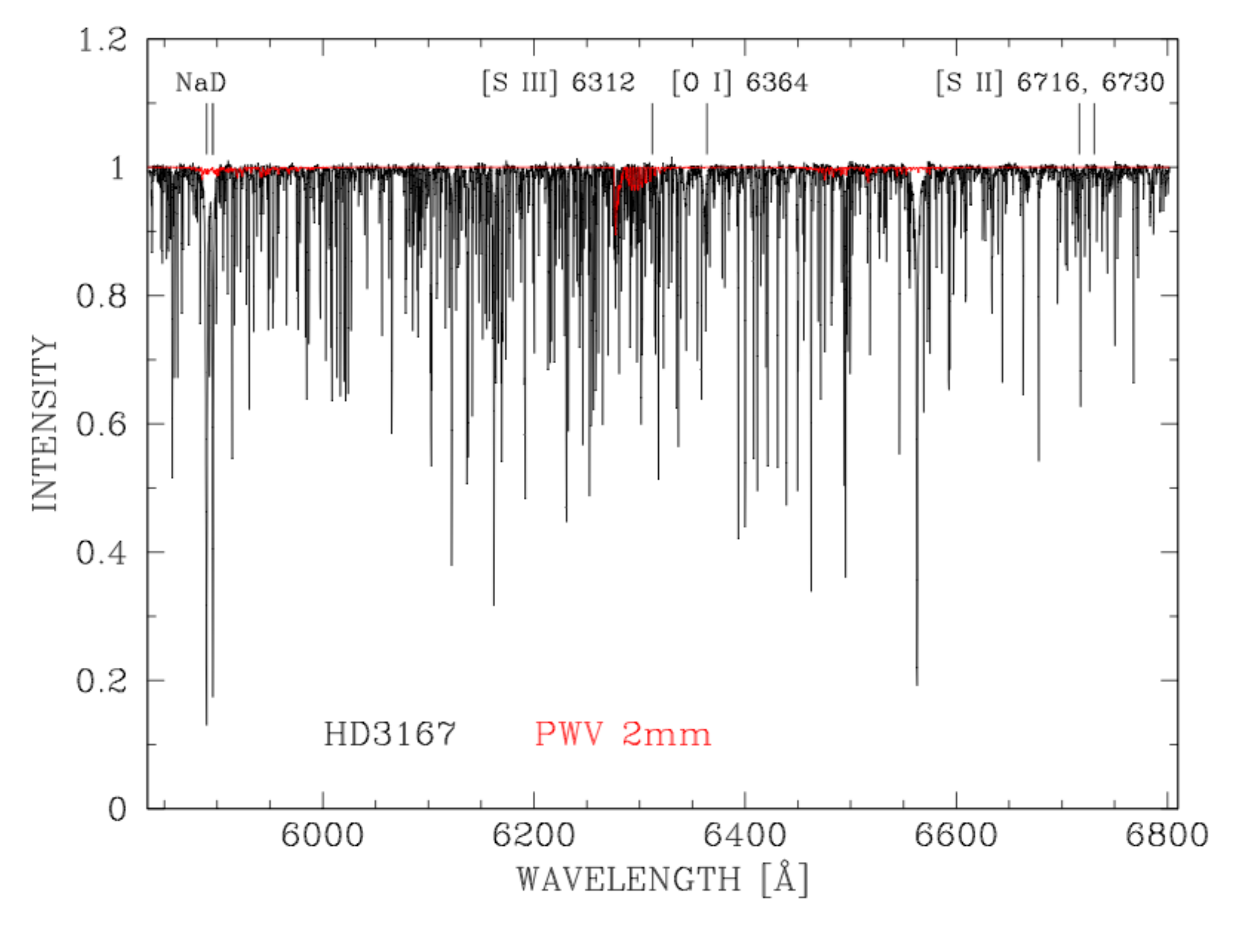}
\caption{Normalized spectrum of HD3167 in the red channel (red line) together with the
telluric spectrum (black line) as calculated by  molectfit for precipitable water 
vapour (PWV) of 2.0 mm.}
\label{PMV}
\end{figure}

\begin{figure}
\includegraphics[height=0.25\textheight,angle=0.0]{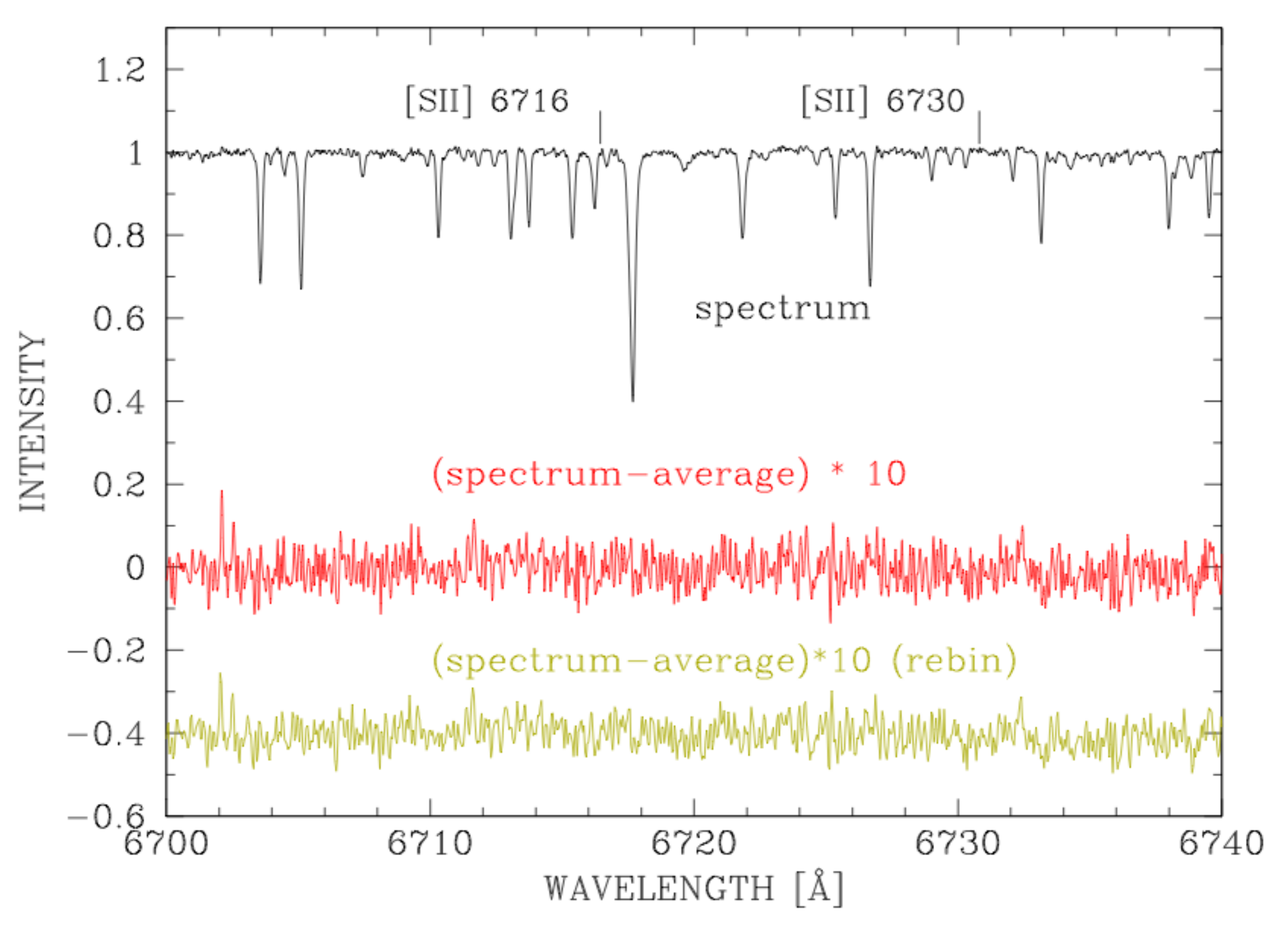}
\caption{Step of the reduction of the frame taken at 03:07:22. The
  black line is the observed spectrum before subtracting the average
  stellar spectrum. The red line is the magnified spectrum
  after the subtraction.  The yellow line is the same but rebinned to
  the resolution of the spectrograph.}
\label{datareduction}
\end{figure}

\begin{itemize}

\item[i.)] UVES is a slit spectrograph. The Thorium-Argon hollow
  cathode lamp used for the initial wavelength calibration is placed
  right in front of the spectrograph and not at infinity.  The
  initial wavelength-calibration, which is based on the Thorium-Argon
  lamps, thus has a small shift with respect to the true wavelength. The
  first step thus was to determine this instrumental shift using the
  telluric lines, and to correct each spectrum for it.

\item[ii.)] The next step was to model the telluric lines. This was
  done using the ESO molecfit tool \citep{noll19}.  The strength of
  the telluric lines varies with the airmass and the weather
  conditions. For example, the $H_2O$ lines become stronger if the
  humidity, or the airmass increases during the observations.  The
  airmass was 1.169 at the beginning of the observations and 1.754 at
  the end.  The precipitable water vapor (PWV) was 2.0 mm until HJD
  2458077.60, and then decreased down to 1.3 mm at HJD
  2458077.64. After that the level increased again and reached 1.6 mm
  at the end of the observations.  On average the PWV was only
  $1.76\pm0.28$ mm.  Fig.~\ref{PMV} shows as example the normalized
  spectrum of HD3167 (black line) together with telluric spectrum
  calculated with molecfit for PWV of 2.0 mm. The model also contains
  all other lines of the Earth atmosphere. Due to the dry conditions
  the $H_2O$ lines were weak and it was thus easy to remove them.

\item[iii.)] The next step was to flux-calibrate the spectra using the
  known spectral energy distribution of the star. We calculated the
  flux of the star at the distance of the Earth in $\rm W\,m^{-2}$ per
  wavelength interval, as well as the total flux emitted by the star
  in Watt per wavelength interval.

\item[iv.)] Given that we expected that any lines from the planet were
  a hundred, or perhaps a thousand times weaker than the stellar
  lines, the subtraction of the stellar spectrum was an important
  step. We tried out several different methods. We subtracted and also
  divided the observed spectrum by the average spectrum. In the
  following sections, we present the results after subtracting the
  averaged stellar spectrum not the ratio of the two spectra. The
  reason is that the resulting spectrum then has the unit $\rm
  W\,m^{-2}$ (per wavelength interval). If we would use the ratio, the
  spectrum would be dimensionless.  However, before we can subtract
  the average stellar spectrum, we have to find out if it varies
  during the observations. A stellar spectrum can change if the
  activity-level of star changes. This is for instance the case if a
  stellar spot rotates into view, or if new spots appear on the
  stellar surface. Since HD3167 has a rotation period of
  $23.52\pm2.87$ days, it rotated only by 2.5 degrees during the
  observations. It is thus unlikely that a new spot rotated into
  view. The lifetimes of spots for stars with similar rotation periods
  are about 100 days \citep{namekata19}. It is thus also unlikely that
  an active region appeared on the visible surface during the four
  hours of observations.  Using the Mt. Wilson $\rm S_{HK}$-index, we
  can also determine the variations in the activity level during the
  observations. In this way we find out whether plage regions, which
  are usually associated with spots, appeared on the stellar surface
  during the observations.  A third reason why a stellar spectrum can
  change are flares. Flares also show up in the Ca{\small II}\,H\&K
  lines. Thus, $\rm S_{HK}$-index shows us whether plages, or spots
  appeared or disappeared on the stellar surface, and whether there
  were any flares. HD3167 is a very inactive star. \cite{vanderburg16}
  determined a Mt. Wilson $\rm S_{HK}$-index of $0.178 \pm 0.005$ and
  the Mt. Wilson $\rm log\,R_{HK}$-index of $-4.97 \pm 0.02$. We found
  a $\rm S_{HK}$-index of $0.1779\pm0.0011$ out-of transit and
  $0.1782\pm0.0007$ \footnote[1]{Relative error of the measurement,
    not the absolute value of the index.}  during transit.  The
  activity level thus did not change significantly during the
  observations and, therefore, it is justified to average the out-of transit
  spectra of the star.  Fig.~\ref{datareduction} shows an individual
  spectrum before and after the subtraction of the average spectrum of
  the host star. This spectrum was obtained at UT 03:07, or roughly at
  the middle of the transit. The figure shows the region containing
  the [SII] lines at 6716.440, and 6730.816 \AA\, (positions marked). We
  also show the spectrum rebinned to the resolution of the
  spectrograph.

\item[v.)] Before we can add up the spectra taken in- and out-of
  transit, we have to shift them according to the RV of
  the planet. Using the values published by \cite{gandolfi17}, we
  calculate that the RV of the planet changes during transit from
  $-43.7\pm3.3$ to $+43.7\pm3.3$ $\rm kms^{-1}$ with respect to the
  star. If we use the values from \cite{christiansen17}, it is
  $-44.5\pm3.4$ to $+44.5\pm3.4$ $\rm kms^{-1}$.  The difference
  between the two is thus smaller than the errors given in the
  publications. On average, the RV of the planet changes by 3.3 $\rm
  kms^{-1}$ from spectrum to spectrum with respect to the star
  \footnote[2]{When planing the observation, we
    calculated how long the exposure time could be, before the
    spectral-lines of the planet would be smeared out by one 
    the resolution element of the spectrograph.}.

\item[vi.)] After computing the weighted average of the spectra taken
  in- and out-of transit, we cut out the regions that we were
  interested in. These are the regions centered on the diagnostic
  lines mentioned in Section~\ref{sectI}, and listed in
  Table~\ref{tab01}.

\item[vii.)] The final step was to measure the flux of the lines using
  IRAF-tool {\it sbands}. We used an extraction width of 10.3 $\rm
  kms^{-1}$ for the lines. The reason for this width is that we have
  to take the combined effects of finite resolution of the
  spectrograph (5.5 $\rm kms^{-1}$), the shift of the planet during
  the exposure (3.3 $\rm kms^{-1}$), and the error of the RV-amplitude
  (typically less than 3.4 $\rm kms^{-1}$), plus the errors of these
  values into account. The length of the continuum strips is always
  100 $\rm kms^{-1}$.  After integrating along wavelength, we obtain
  the fluxes of the lines in $\rm W\,m^{-2}$, and W, respectively. The
  results are given in Table~\ref{tab01}.

\end{itemize}

\section{Results}
\label{sectIII}

\begin{figure}
\includegraphics[height=0.25\textheight,angle=0.0]{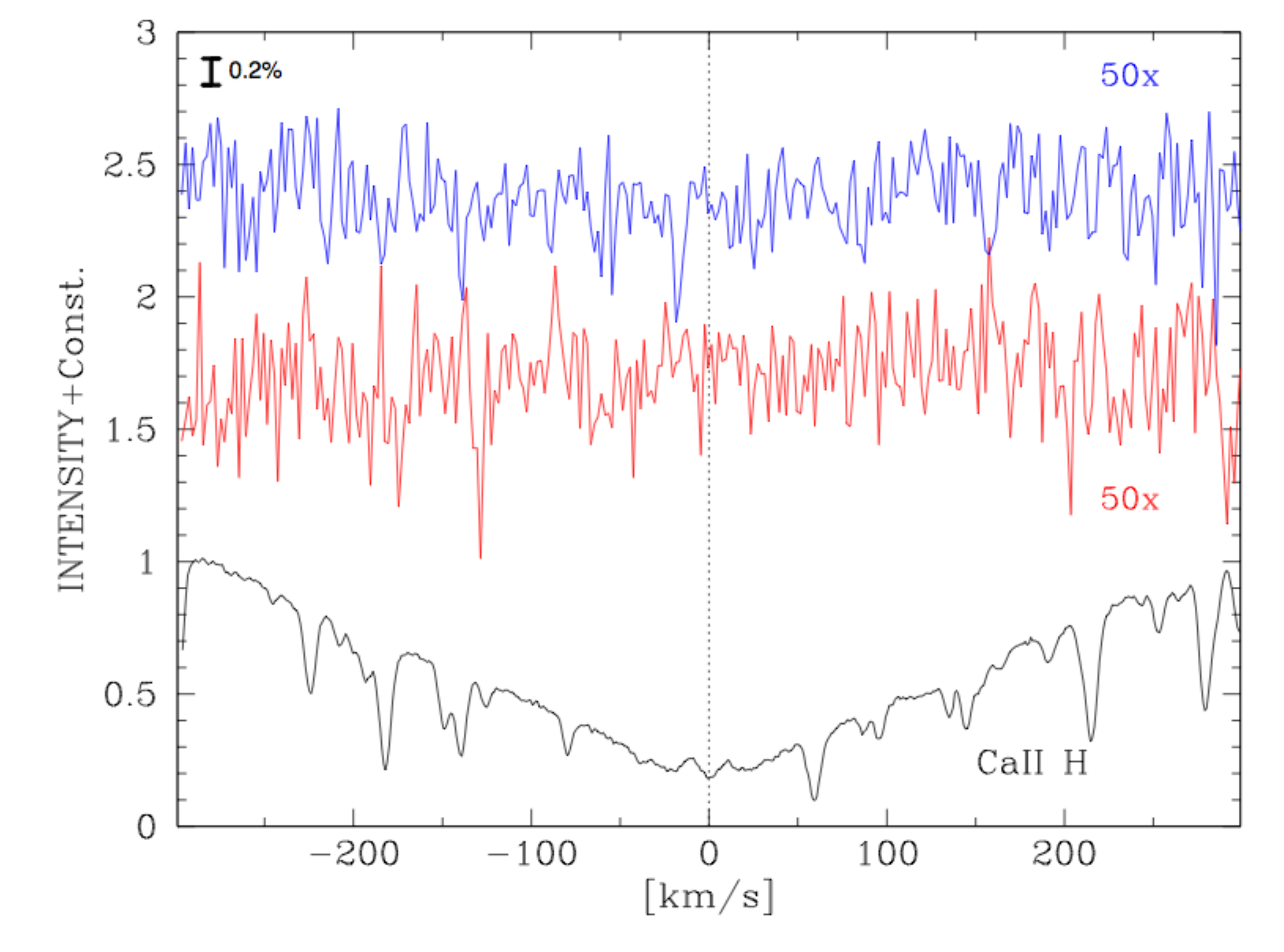}
\caption{Average residual spectrum in the Ca{\small II}\,H line after
  subtracting the average out-of transit spectrum of the star.  As
  explained in the text, the red line is the residual of the
  in-transit spectrum, in other words, the spectrum of the planet.
  The blue line is the residual of the out-of transit spectrum.  We
  magnified the residuals by a factor of 50 so that the signal can be seen.
  The scale for the red and blue lines is shown at upper left corner.
  The black line is the stellar spectrum.}
\label{CaIIH}

\end{figure}

\begin{figure}
\includegraphics[height=0.25\textheight,angle=0.0]{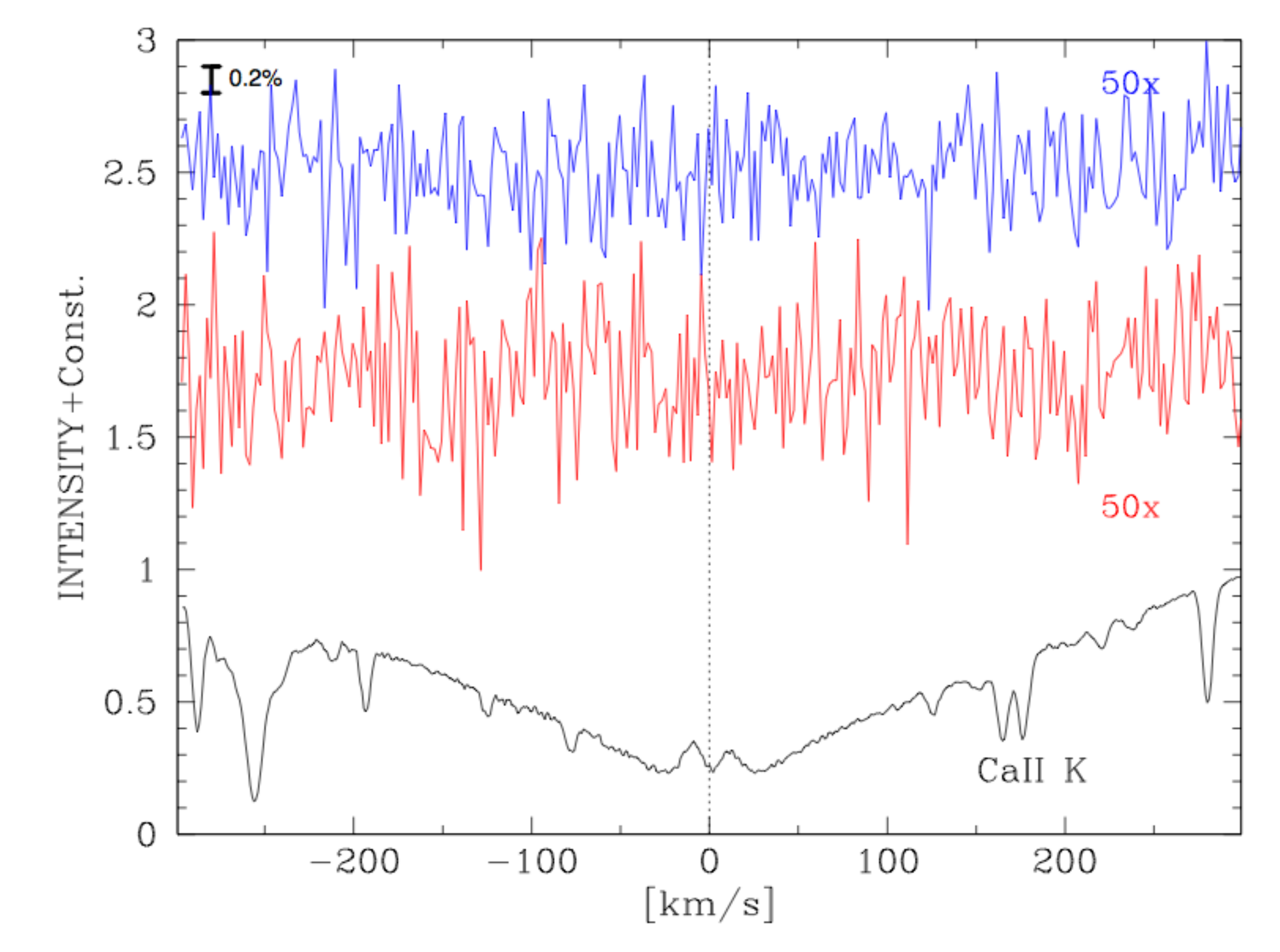}
\caption{Same as Fig.~\ref{CaIIH} but for Ca{\small II}\,K line.}
\label{CaIIK}
\end{figure}

\begin{figure}
\includegraphics[height=0.25\textheight,angle=0.0]{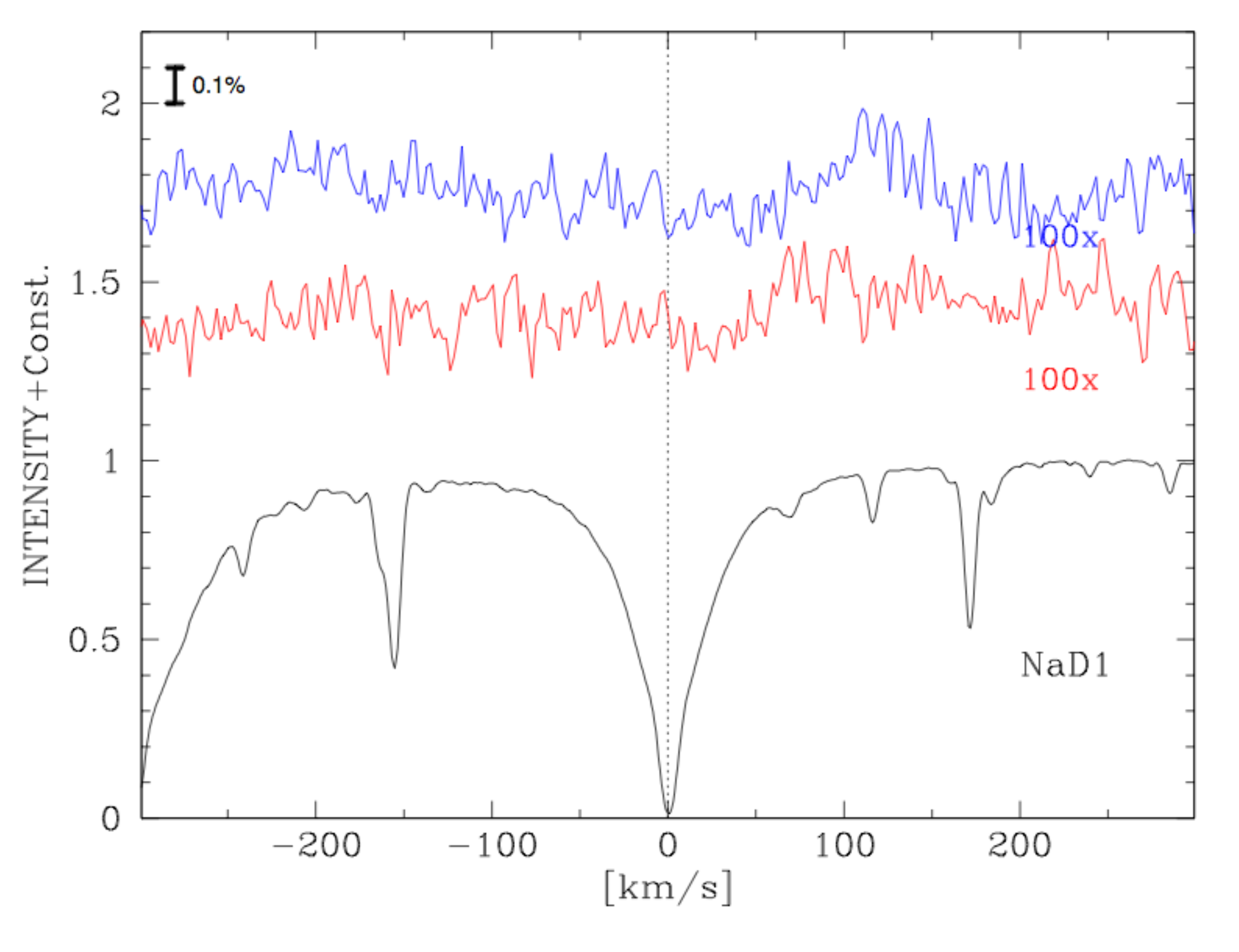}
\caption{Same as Fig.~\ref{CaIIH} but for $\rm NaD_1$ line and the
  residuals magnified by a factor of 100. One unit thus corresponds to
  $10^{-2}$ of the continuum level of the star.}
\label{HD3167NaD1}
\end{figure}

\begin{figure}
\includegraphics[height=0.25\textheight,angle=0.0]{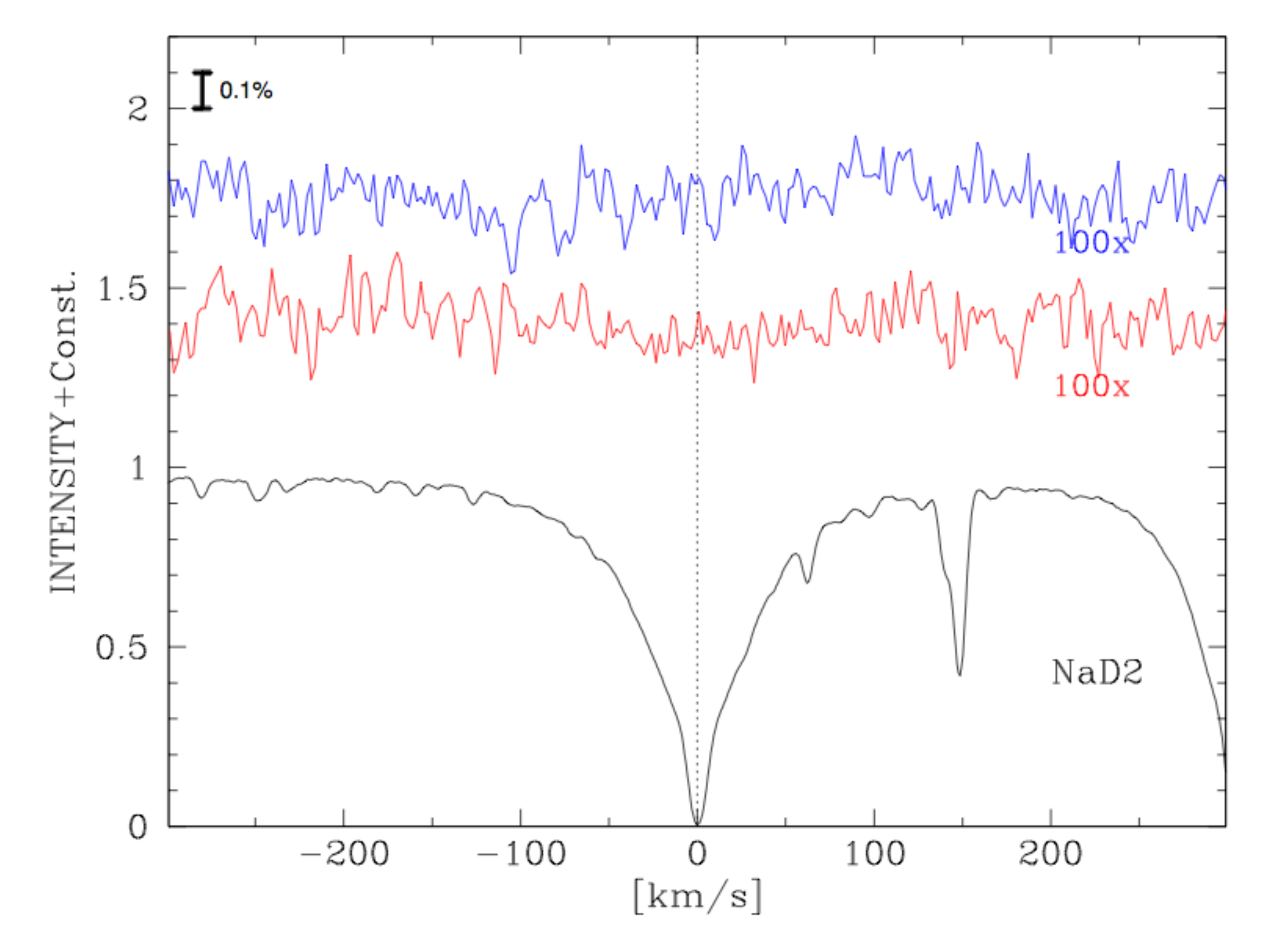}
\caption{Same as Fig.~\ref{HD3167NaD1} but for $\rm NaD_2$ line.}
\label{HD3167NaD2}
\end{figure}

\begin{figure}
\includegraphics[height=0.25\textheight,angle=0.0]{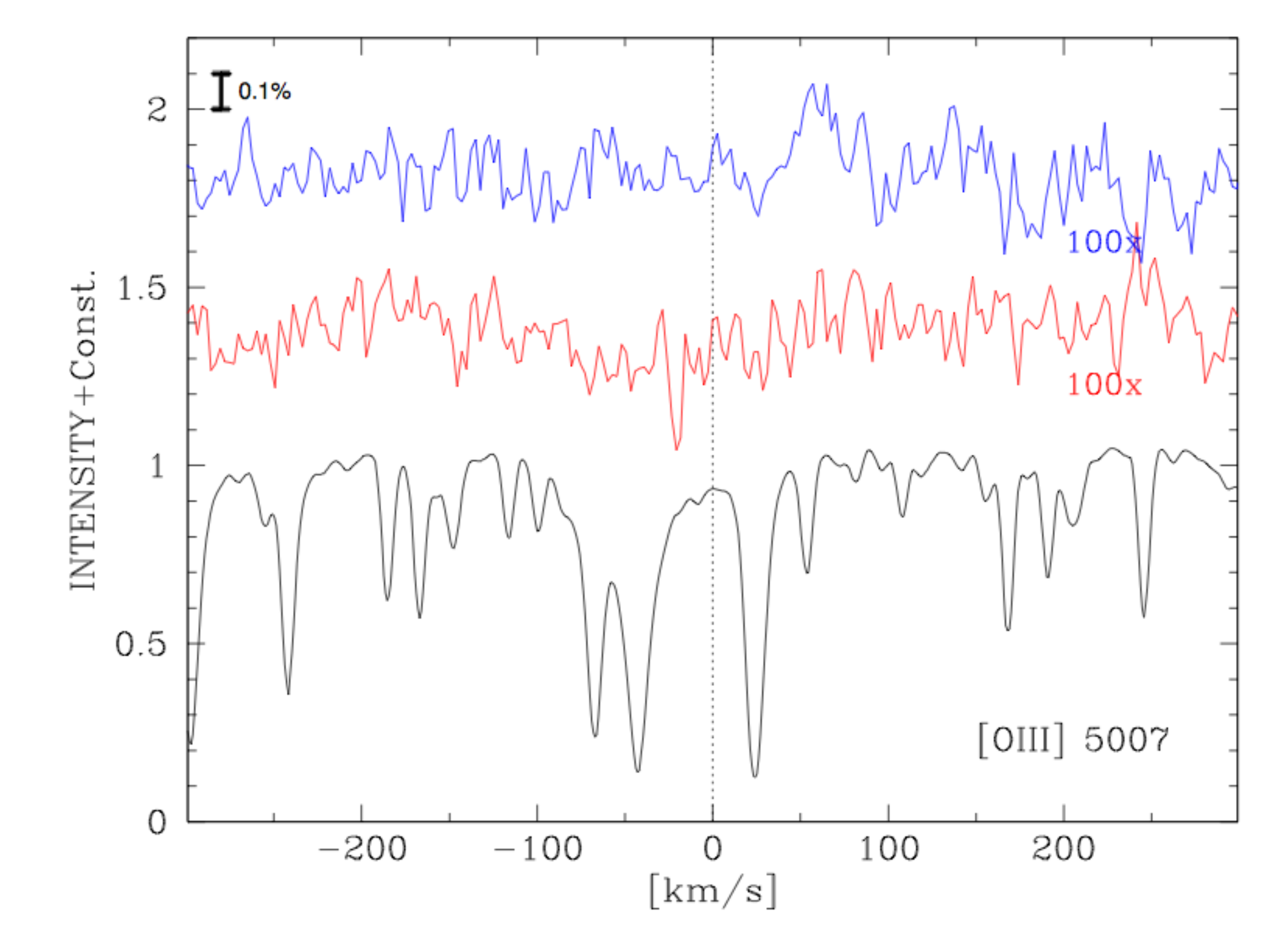}
\caption{Same as Fig.~\ref{HD3167NaD1} but for [OIII] 5007 line.}
\label{HD3167o5007}
\end{figure}

\begin{figure}
\includegraphics[height=0.25\textheight,angle=0.0]{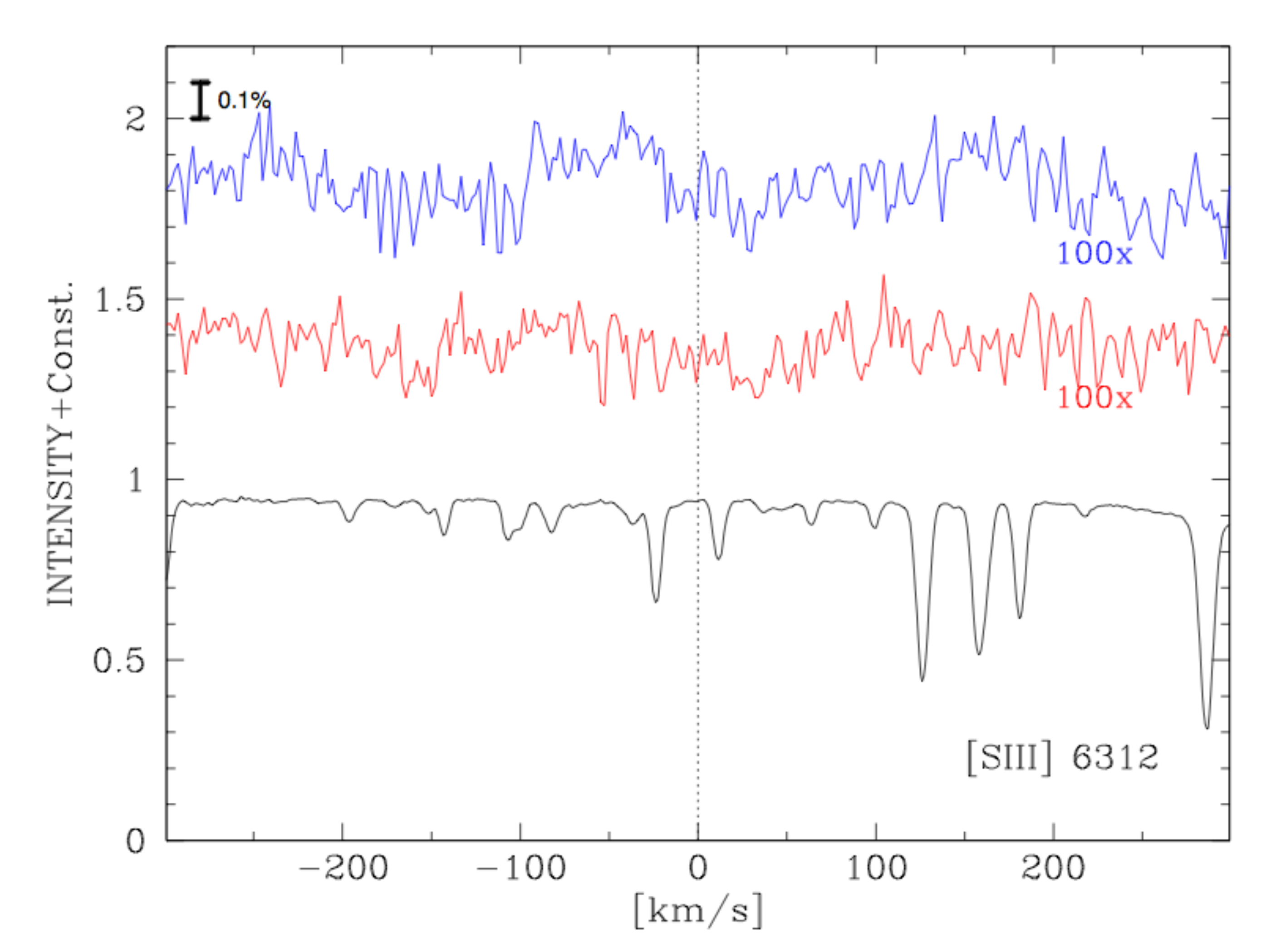}
\caption{Same as Fig.~\ref{HD3167NaD1} but for the [SIII] 6312 line.}
\label{HD3167SIII6312}
\end{figure}

\begin{figure}
\includegraphics[height=0.25\textheight,angle=0.0]{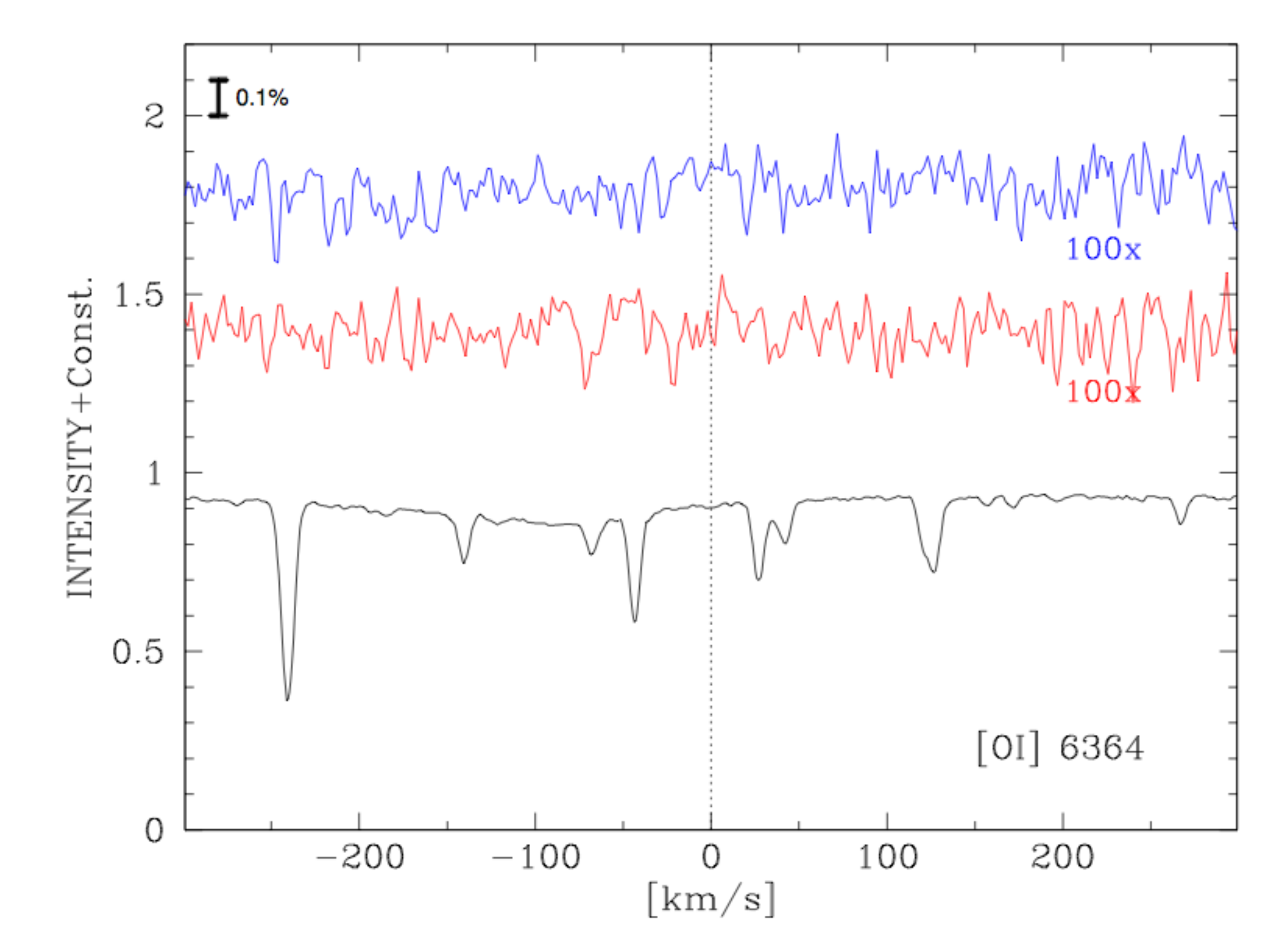}
\caption{Same as Fig.~\ref{HD3167NaD1} but for the [OI] 6364 line.}
\label{HD3167OI6364}
\end{figure}

\begin{figure}
\includegraphics[height=0.25\textheight,angle=0.0]{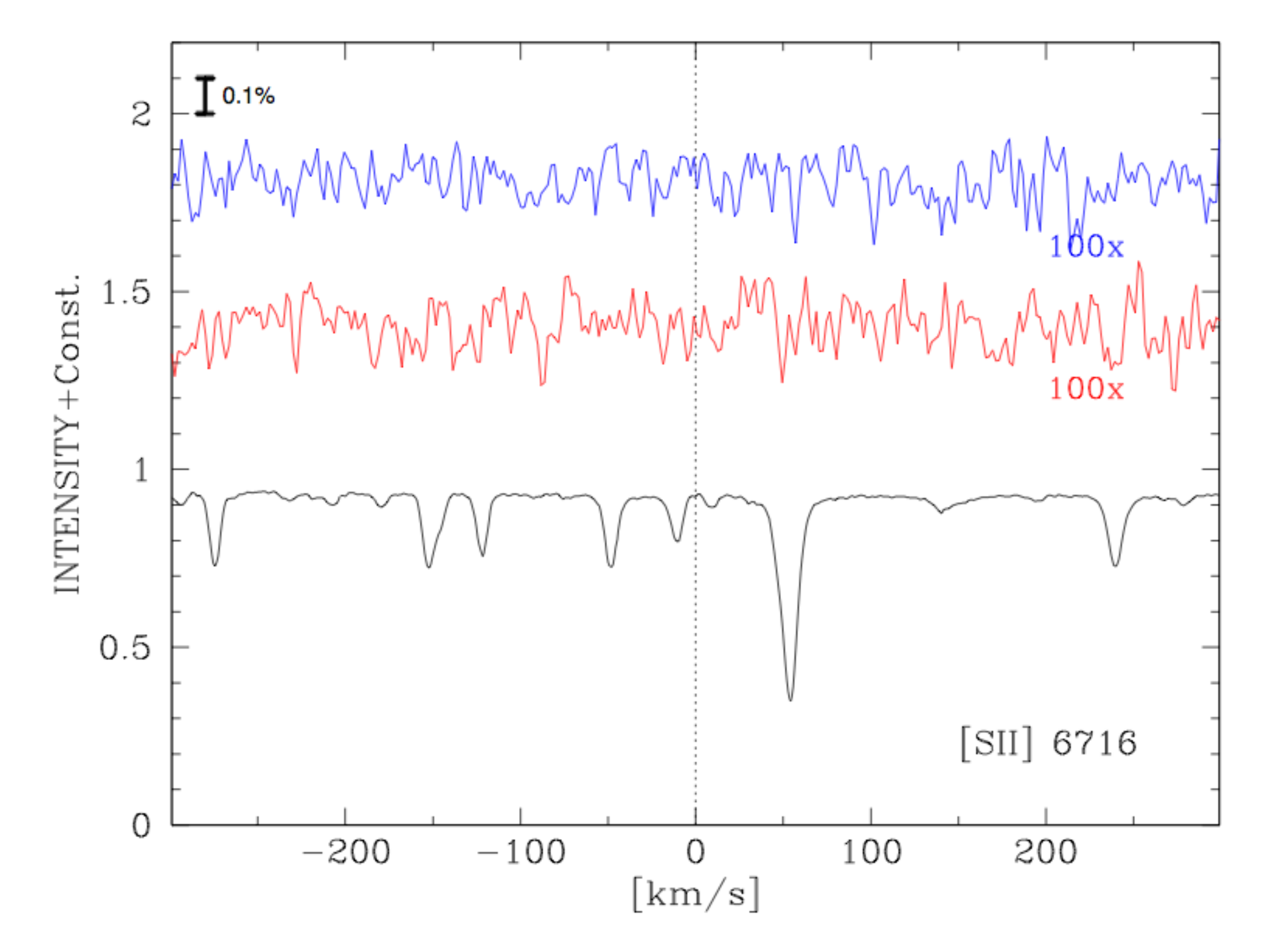}
\caption{Same as Fig.~\ref{HD3167NaD1} but for [SII] 6716 line.}
\label{HD3167SII6716}
\end{figure}

\begin{figure}
\includegraphics[height=0.25\textheight,angle=0.0]{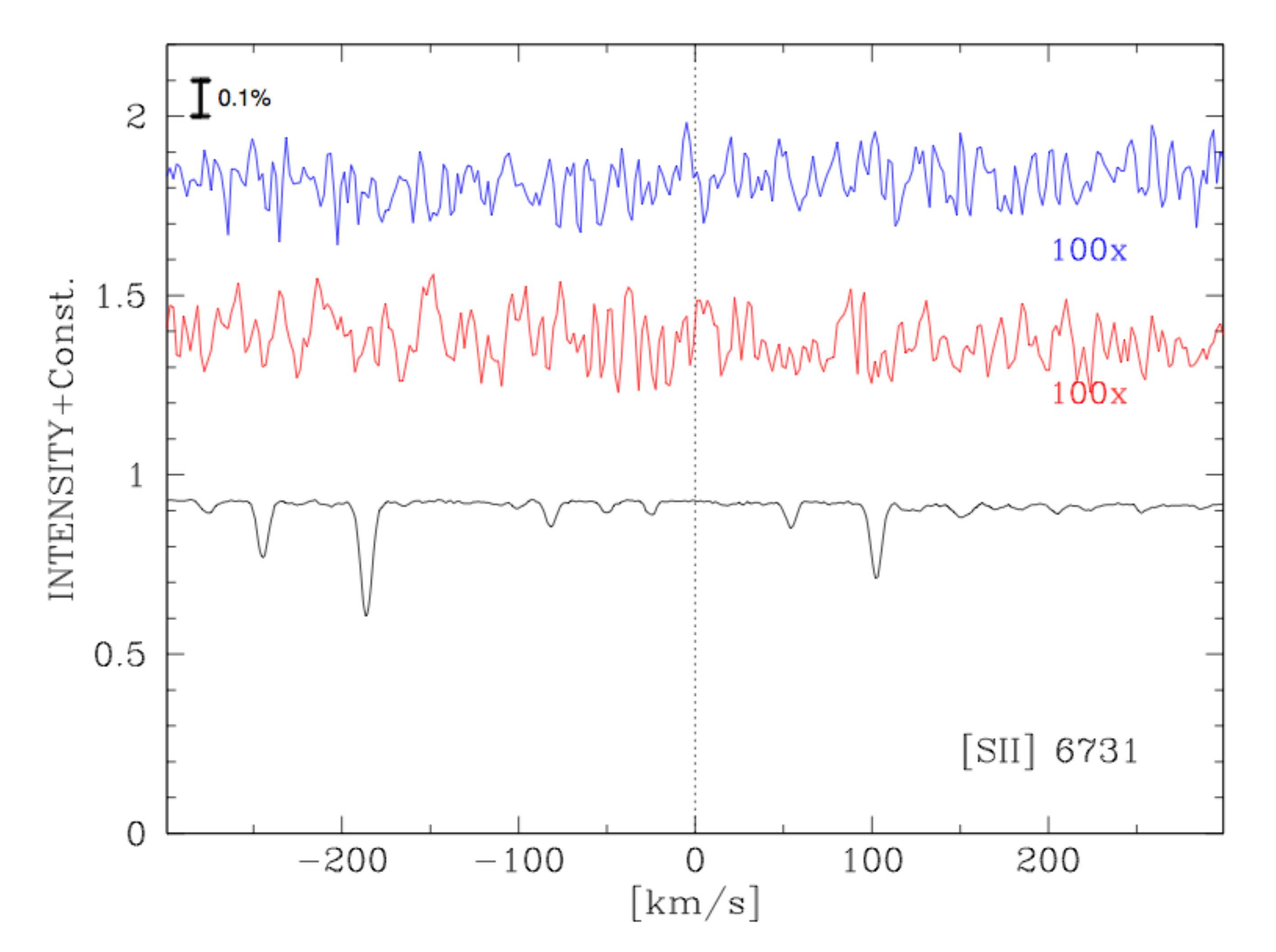}
\caption{Same as Fig.~\ref{HD3167NaD1} but for the [SII] 6731 line. }
\label{HD3167SII6731}
\end{figure}

The results are presented in Figs.~\ref{CaIIH} to \ref{HD3167SII6731}.
Each figure presents a spectrum for one of the spectral lines listed in
Table~\ref{tab01}.  The red lines in these figures are the residuals
of the in-transit spectra after subtracting the stellar and telluric
spectra. In other words, the red-lines are the spectra of the planet.
  
The blue lines are the residuals of the out-of transit spectra. They
should not contain any spectral lines of planetary origin.  We show these
spectra as a reference for the noise level of the residual spectra.
  
We use the units of $\rm kms^{-1}$ for the wavelength rather than \AA
, because the resolution of an \'Echelle spectrograph is almost
constant in velocity over the whole observed wavelength range (5.5
$\rm kms^{-1}$), and also because it makes it easy to see were the
line from the planet should be. The line originating from the planet
would be centered at zero velocity and would have a width of 10.3 $\rm
kms^{-1}$, given the resolution of the spectrograph, the RV-shift of
the planet during the exposure and taking the error of the
RV-amplitude into account.
  
Since the spectra are flux-calibrated, in principle  the unit on the 
Y-axis is Watt per wavelength-interval. However, since we like to
show the residual spectra together with the spectrum of the star, we
have to magnify the residual spectra by a factor of 100 (a factor of
50 for the Ca{\small II}\,H,K lines). Thus, one unit of the residual
spectra corresponds to $10^{-2}$ ($2 \times 10^{-2}$ for the Ca{\small
  II}\,H,K lines) of the continuum flux of the star.  In order to show
the different spectra more clearly, we moved the red and the blue lines
upwards so that the spectra are well separated.  The scale of the
residual spectra is shown at the upper left corner in each figure.
  
We also show the stellar spectrum (black line) in Figs.~\ref{CaIIH} to
Fig.~\ref{HD3167SII6731} to demonstrate the location of the stellar
lines.  The noise level is higher at those wavelengths where the star has
deep lines. However, since the spectrum of the star moved from
$-44.5\pm3.4$ to $+44.5\pm3.4$ $\rm kms^{-1}$ with respect to the
spectrum of the planet, the effect from the deep stellar lines was
significantly reduced.

We measured the flux in a region with a width of 10.3 $\rm kms^{-1}$
at the position of each spectral line.  The upper limits for the
fluxes of the lines emission are given in Table~\ref{tab01}, and the
system's parameters in Table~\ref{tab02}. The first two columns give
the identification of the line and the wavelength. The third column
shows the the line fluxes from the planet relative to the flux of the
star at that line ($I_{p,\lambda}/I_{*,\lambda}$).  The significance
of the result is given in the fourth columns. Since all values are
below 3$\sigma$, none of the lines were detected and we consider the
values in the fourth column as upper limits.  The upper limits that we
derive are approximately $1.5 \times 10^{-3}$ for the Ca{\small
II}\,H,K lines and $7.2\times 10^{-4}$ and $3.3\times 10^{-4}$ and for
the $\rm NaD_{1,2}$ lines, respectively.  For comparison,
\cite{ridden16} detected the Ca{\small II}\,H,K in 55 Canc\,e at a
level of $I_{p,\lambda}/I_{*,\lambda}= 7 \times 10^{-2}$ and
$I_{p,\lambda}/I_{*,\lambda} = 2.3 \times 10^{-3}$ the $\rm NaD_{1,2}$
lines, respectively. Thus, our upper limits are even lower than the
$I_{p,\lambda}/I_{*,\lambda}$-values for 55\,Cnc\,e.

\begin{table*}
\caption{Upper limits of the line fluxes from the planet }
\begin{tabular}{l l l c c c l}
\hline
line     & $\lambda $ & $I_{p,\lambda}/I_{*,\lambda}$ & $significance^1$ & observed [$W m^{-2}$] & total emission [W] & remarks$^2$  \\
           &       [\AA]     &                               & $\sigma$  &  $3 \sigma$ upper limits  & $3 \sigma$ upper limits & \\
\hline
Ca{\small II}\,K & 3933.666 & $-0.00153\pm 0.00085$ & 1.8 & $<4.5\times 10^{-20}$ & $<1.2\times 10^{18}$ & 55\,Cnc e \\
Ca{\small II}\,H & 3968.468 & $-0.00156\pm 0.00097$ & 1.6 & $<5.6\times 10^{-20}$ & $<1.5\times 10^{18}$ & 55\,Cnc e \\
$\rm [S II]$  & 4068.600 & $-0.0039\pm0.0020 $ & 2.0 & $<7.3\times 10^{-19}$ & $<1.9\times 10^{19}$ & Io \\
$\rm [S II]$  & 4076.349 & $-0.0003\pm0.0020$  & 0.2 & $<7.5\times 10^{-19}$ & $<2.0\times 10^{19}$ & Io \\
Ca{\small I}  & 4226.728 & $-0.024\pm0.024$     & 1.0 & $<3.4\times 10^{-18}$ & $<9.2\times 10^{19}$ & Io, Mercury \\
$\rm [O III]$ & 4356.794 & $-0.0036\pm0.0020$   & 1.8 & $<9.2\times 10^{-19}$ & $<2.5\times 10^{19}$& \\
$\rm [O III]$ & 5006.843 & $-0.0001\pm0.00045$  & 0.2 & $<2.4\times 10^{-19}$ & $<6.4\times 10^{18}$ & Io \\
$\rm NaD_2$   & 5889.953 & $\,\,\,\,\,0.00033\pm0.00040$ & 0.8 & $<2.4\times 10^{-19}$ & $<6.5\times 10^{18}$ & Io, 55\,Cnc\,e, Mercury \\
$\rm NaD_1$   & 5895.923 & $\,\,\,\,\,0.00072\pm0.00027$ & 2.6 & $<1.7\times 10^{-19}$ & $<4.5\times 10^{18}$ & Io, 55\,Cnc\,e, Mercury \\
$\rm [S III]$ & 6312.06  & $-0.00003\pm0.00046$          & 0.1 & $<2.7\times 10^{-19}$ & $<7.3\times 10^{18}$ & Io \\
$\rm [O I]$   & 6363.776 & $\,\,\,\,\,0.00033\pm0.00029$ & 1.1 & $<1.7\times 10^{-19}$ & $<4.6\times 10^{18}$ & \\
$\rm [S II]$  & 6716.440 & $\,\,\,\,\,0.00040\pm0.00029$ & 1.4 & $<1.8\times 10^{-19}$ & $<4.8\times 10^{18}$ & Io \\
$\rm [S II]$  & 6730.816 & $\,\,\,\,\,0.00040\pm0.00032$ & 1.2 & $<2.0\times 10^{-19}$ & $<5.4\times 10^{18}$ & Io \\
\hline
\end{tabular}
\label{tab01}
 \\
$^1$ All excesses below 3$\sigma$ are considered insignificant and
 taken as upper limits.\\
$^2$ Examples for objects were this line has been detected. \\
$I_{p,\lambda}/I_{*,\lambda}$ is the intensity of the planet in respect to the
star at the wavelength of the spectral-line.
 \end{table*}

\begin{table}
\caption{Parameters of the star HD\,3167 and planet HD\,3167b taken
  from \citep{gandolfi17}.}
  \begin{tabular}{l l }
\hline
Parameter    & Value \\
\hline
Stellar mass [M$_\odot$] & 0.877 $\pm$ 0.024 \\
Stellar radius [R$_\odot$] & 0.835 $\pm$ 0.026 \\
Stellar age [Gyr] & 5.0 $\pm$ 4.0   \\
Stellar equilibrium temperature [K]  & 5286 $\pm$ 40  \\
Stellar rotational period [days] & 23.52 $\pm$ 2.87 \\
Distance to the system [pc] & 45.8 $\pm$ 2.2 \\
Planetary mass [M$_\oplus$]   & 5.69 $\pm$ 0.44  \\
Planetary radius [R$_\oplus$]   & 1.574 $\pm$ 0.054 \\
Planetary equilibrium temperature [K]  & 1759 $\pm$ 20  \\
Semi-major axis [au]  & 0.01752 $\pm$ 0.00063  \\
Inclination [degree] & 88.6$^{+1.0}_{-1.5}$ \\
\hline
\end{tabular}
\label{tab02}
\end{table}

\section{Possible origin of the atmosphere}
\label{sectIV}

There are several mechanisms to produce an atmosphere on a close-in
rocky exoplanet. First, the atmosphere can be generated by volcanic
activity driven by internal heating sources, and second, if the
temperature on the surface is high enough, the rocks on the surface
can melt thus creating a mineral atmosphere. In this section, we
discuss both mechanisms of atmosphere formation.

Volcanic activity of a planet requires an internal heating source as a
driver. There are several sources of internal energy known for
planets: heat is produced due to radioactive decay, mantle
differentiation, core and inner core formation, tidal heating, and
induction heating. The first four of these mechanisms are the dominant
ones in the Earth. They can be very important for young planets, but
the old age of HD3167\,b indicates that these heating sources are
likely not the dominant ones at present \citep{noack16}. The latter
two mechanisms, while insignificant for the Earth, can be very
powerful in planets orbiting close to their host stars. From the
observations of Io we know that a powerful internal heat source can
drive enormous volcanic activity \citep{peale79}. Tidal heating, to be
substantial, requires an eccentric or inclined orbit of a
planet. Since the orbit of HD3167\,b is likely highly inclined, one
can expect this heating source to be quite important for HD3167b
\citep{beuthe13}. However, due to a relatively high mass of HD3167\,b, volcanic
outgassing from this planet is likely inefficient
\citep{noack17,dorn18}. \cite{dorn18} have shown that at small masses
(below 2-3 M$_\oplus$) outgassing positively correlates with the
planet's mass, since it is controlled by volume of the mantle, but at
higher masses outgassing decreases with the planet's mass, which is
due to the increasing pressure gradient that limits melting to
shallower depths. Since tidal heating produces heat mostly deep in the
planetary mantle, it can not melt the mantle of a planet as massive as
HD3167\,b. Therefore, we can conclude that even though one can expect
strong tidal heating in the interiors of HD3167\,b, it can unlikely drive
strong volcanic activity on this planet.

Another newly identified heating source for exoplanets is induction
heating \citep{kislyakova17}. Electromagnetic induction is the
production of voltage across an electrical conductor if the magnetic
field around it varies in time. This voltage, or electromotive
force, generates alternating currents flowing inside the body, which
then dissipate and produce heat inside the conductor. This heating
source requires a presence of a strong varying magnetic field around
the planet. For HD3167\,b, the source of the variation of the magnetic
field around the planet is the planetary motion on its highly inclined
polar orbit. \cite{kislyakova18} have shown that induction heating can
be very substantial for planets on inclined orbits, if the stellar
magnetic field is strong enough. Below we calculate the power of
induction heating inside HD3167\,b. We have followed the procedure
described in \cite{kislyakova18} for a planet on an inclined orbit.

\begin{figure}
\includegraphics[height=0.24\textheight,angle=0.0]{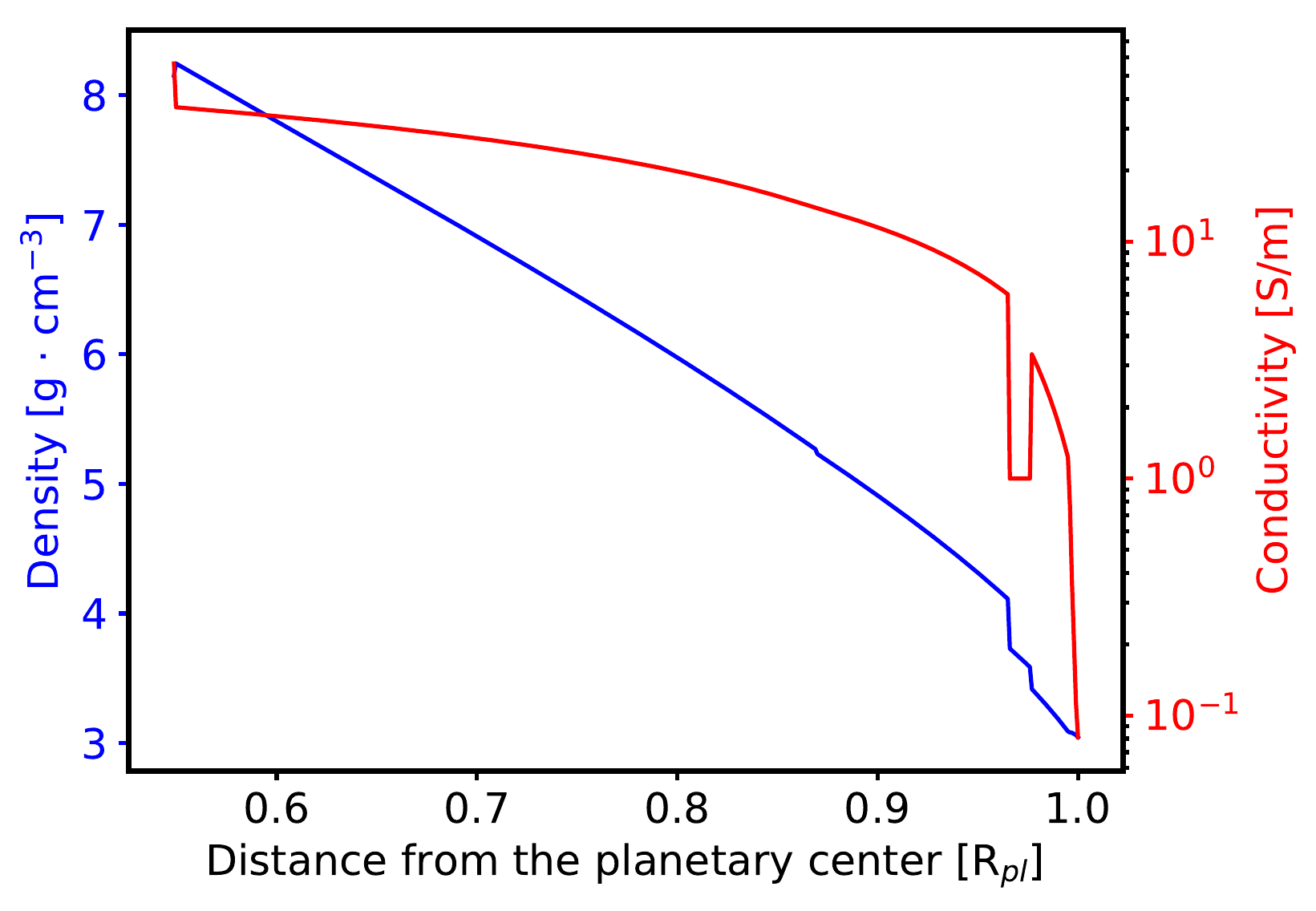}
\caption{Density (blue) and electrical conductivity (red) profiles for
  HD3167\,b calculated with the interior model CHIC \citep{noack16}.}
\label{Profiles}
\end{figure}

First, we have calculated the density and electrical conductivity
profiles for HD3167\,b using the code CHIC \citep{noack16} that are
shown in Fig.~\ref{Profiles}. We have assumed a composition with a
slightly higher iron content than that of the Earth, namely, of
40\%. We further assumed that the iron accumulates in the planet's
core. Electrical conductivity has been calculated for a given
composition of a mineral layer in the mantle and for a calculated
temperature profile. We have assumed the equilibrium temperature of
HD3167\,b at its surface as a boundary condition (see
Table~\ref{tab02}). We have used the procedure to calculate induction
heating in a planet with a conductivity varying with depth, as
described in \cite{kislyakova17}. In this model, induction equation is
solved in every layer, and the conductivity values at the surface and
in the planetary core are used as boundary conditions. We also need
the strength of the stellar magnetic field as an input
parameter. Although the magnetic field of HD3167 has not been
measured, one can conclude from its old age of 5~Gyr and a relatively
slow rotational period of 23 days that this star likely has a
relatively low magnetic field. Following \cite{vidotto14} and the
observations of the Sun, which is also an inactive star, we have
assumed the strength of the global dipole component of the stellar
magnetic field of 2-5~G. Although higher harmonics of the stellar
magnetic field can also be present, we have not considered them
here. They are unlikely to produce a strong heating effect due to
their sharp decline with the distance to the star.

\begin{figure}\includegraphics[height=0.24\textheight,angle=0.0]{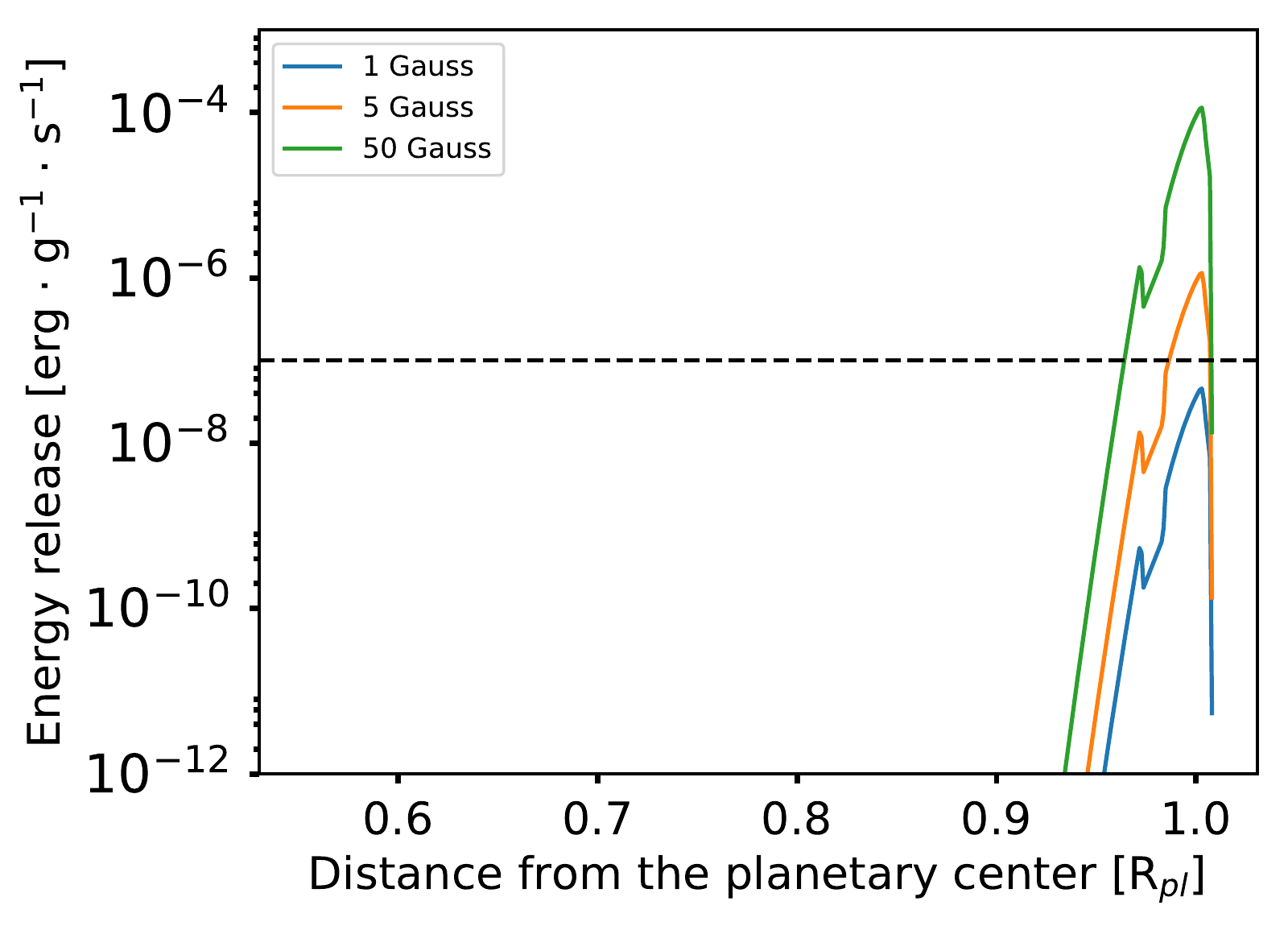}
\caption{Induction heating inside HD3167\,b for three values of the
  stellar magnetic field: 1, 5, and 50~G. The present-day global
  magnetic field of HD~3167 is likely weak and close to 1~G. The
  picture shows the heating rates inside the planet. The dashed line
  indicates an approximate limit necessary to melt the rocks on a
  geological time scale.}
\label{IndHeating}
\end{figure}

The results of our calculation are presented in
Fig.~\ref{IndHeating}. The figure shows internal heating rates inside
HD3167\,b for three values of the stellar dipole field: 1, 5, and
50~G. These cases correspond to the total energy release inside
HD3167\,b of $1.8\times10^{19}$, $4.5\times10^{20}$, and
$4.5\times10^{22}$ erg~s$^{-1}$, respectively. While the first two
values are typical for an old, inactive star, the last value has been
selected to illustrate possible induction heating in the planet when
this stellar system was young. The dashed line indicates an
approximate limit of the internal heating rate necessary to melt the
planetary mantle, if the planet is exposed to this amount of heating
for a geologically long time \citep{kislyakova17}. Note that this limit
can vary within approximately $10^{-6}-10^{-8}$~erg~g$^{-1}$~s$^{-1}$,
with the lower value more typical for small planets. Unlike tidal heating
which leads to energy release deep in the planetary mantle, induction
heating (or skin effect) preferentially heats the very upper part of
the planetary mantle. According to \cite{dorn18}, this indicates that
in this case some melt and, therefore, some volcanic activity could be
produced, because the pressure in the upper mantle is not yet so high
that it would suppress the melt formation. Our non-detection of traces of
volcanic activity indicates that the global magnetic field of HD~3167
is likely weak and of the order of only 1-2~G. We can also conclude
that induction heating must have been stronger in HD3167\,b in the
past when the stellar magnetic field was stronger. At present, the
exosphere of HD3167~b is likely generated by the melting of the
surface rocks by intense stellar radiation.

\section{Discussion}
\label{sectV}

The architecture of the HD3167-system is unusual, which makes the
system an interesting target for many different studies. HD\,3167\,b
is a transiting planet with an orbital period of 0.96 days, and an
inclination of $\rm i=88.6^{+1.0}_{-1.5}$ degrees.  The next outer planet
is supposed to be the non-transiting planet HD~3167\,d.  However, as
we pointed out above, it is not certain that this planet exists. The
third, or perhaps second planet of the system is the transiting planet
HD3167\,c which has an orbital period of 29.8 days and $\rm
i=89.6\pm0.2$ degrees. Recently, \cite{dalal19} found that HD3167\,c is in
a nearly polar orbit.  Since most planetary systems are coplanar,
HD3167\,b should also have a nearly polar orbit.

Because induction heating requires the magnetic field at the position
of the planet to vary, a system with planets in polar orbits is
a good target for studying induction heating \citep{kislyakova18,kislyakova17}. 
HD3167 is also a good target, because it is a nearby, bright star 
that allows to take high precision measurements.

However, induction heating also depends on the strength of the
magnetic field of the star. The rotation period of HD3167 is
$23.52\pm2.87$ days and the activity index $\rm log\,R'_{HK}=-5.04$. For
comparison, the Sun has $\rm log\,R'_{HK}=-4.96$ and a rotation period of
25 days \citep{wright04}.  There are 21 photometric measurements of HD3167 
in the APPLAUSE
\footnote{https://www.plate-archive.org/applause} data-base. The
average brightness of the star is $\rm V=9.03\pm0.10\,mag$. Given that
the average error of the photometry is 0.15 mag, the stars is, within
errors, not variable. This is also consistent with a star of a solar
activity level. The activity level of HD~3167 is basically the same
as the Sun. We thus concluded (Section~\ref{sectIV}) that at
present neither induction nor tidal heating are capable to
produce significant volcanic outgassing on HD3167\,b. This is in
agreement with a non-detection of sulfur lines in our observations,
since sulfur is mostly produced by classical volcanic outgassing.

This leaves the third possibility for the atmosphere production, namely,
a mineral atmosphere, which is produced due to melting of the
surface crust by intense stellar radiation \citep{schaefer09}. The high
equilibrium temperature of HD3167\,b of $\rm T_{eq}=1759\pm20\,K$
suggests that it should indeed have a tenuous mineral atmosphere
\citep{ito15,miguel11}, likely dominated by sodium and oxygen
\citep{miguel11}.

How meaningful are the upper limits derived? Apart from the Earth and
Venus, where traces of ongoing volcanic activity have recently been
identified \citep{shalygin15}, Io is the only other object in the solar
system that has active volcanism. Sulfur ejected from Io forms a
plasma ring around Jupiter, known as the Io plasma torus. Our
non-detection of volcanic activity on HD3167~b indicates that the
magnetic field of HD~3167 is likely weak, which is consistent with the
relatively slow rotation and the low X-ray brightness of the star.

On Io, \cite{morgan82} have detected fluxes of $215\pm28$ Rayleigh for
the [SII] 6731 \AA \ which corresponds to a total emission of $5\times
10^{15}$~W. Our upper limit for that line is $5.4\times 10^{18}$~W
and thus a factor of 1000 higher. However, we should keep in mind that
HD3167\,b has 5.5 times the radius and 30 times the surface area of
Io, and that the internal and external heating of HD3167\,b is supposed to
be correspondingly stronger. It is thus reasonable to assume that the lines could
also be much stronger. The non-detection of a 
volcanic atmosphere caused by tidal heating agrees with the results by
\cite{noack17} and \cite{dorn18} that massive planets do not exhibit
efficient outgassing. \cite{demory16} have also argued that the hight of
volcanic plumes can not be much higher than about 10 scale heights in
the atmosphere, corresponding to a few hundred km in the case of the a super
Earth. If this is true, it would be very difficult to detect a volcanic atmosphere 
in any super Earth.

We can also follow \cite{ridden16} and calculate the size of the
absorption region and compare it to the size of the Hill-sphere
for an optically thick line. For the $\rm Ca^+$ exosphere in
55\,Cnc\,e they found that it is approximately five times larger than
the Roche lobe radius, or about 8 $\rm R_{Hill}$. For NaD they found
roughly a Roche lobe radius, or 1.5 $\rm R_{Hill}$. The 3$\sigma$
upper limits correspond to a regions of 0.4-0.5 Hill spheres for the
Ca{\small II}\,H,K lines and 0.1-0.3 Hills spheres the $\rm NaD$
lines. Our upper limits are thus smaller than size of these regions in
55\,Cnc\,e.

We can also compare our upper limits of the line-fluxes with the total
amount of radiation that the planet receives, because radiative
heating from the star should be one of the main energy sources of the
volcanism. The upper limits for the lines are typically 500-1000 times
smaller than the radiation that the planet receives from the star
which is $2.9\times10^{21}$\,W.

In our observations, we have searched for both sodium and oxygen
lines, but we were only able to obtain upper limits for the signal. We
argue that because of the relatively high mass, and the relatively low
magnetic field strength, detecting the signature of volcanism is very
difficult. This is consistent with out measurements. An ideal target
would be a lower mass USP orbiting an active star with a stronger
magnetic field, or a lower mass USP on an eccentric or inclined orbit.
As we have shown, an object like HD3126 but with a magnetic field strength
of 50 Gauss would be perfect. The prospects for detecting volcanism
are better for planets of M-stars because these tend to have higher
magnetic field strength than solar-like stars.

The best objects to detect mineral atmospheres are planets with high
equilibrium temperatures, likely higher than that of HD3126\,b.
However, HD3167\,b still is an interesting target for detection of a mineral
atmosphere.  We note that in the case of 55\,Cnc\,e, the Ca{\small
  II}\,H\&K line were detected in only one of the five transit
observed. If the signal is that variable as in 55\,Cnc\,e it might
well be that these lines are also detected in HD3167\,b if the
observations are just repeated several times.

\section{Conclusions}
\label{sectVI}

HD3167\,b is one of the best targets to search for volcanic activity
and a Mercury-like exosphere of a rocky planet.  It is not only a
planet that orbits at a very short distance from the star, also it host
star is only at a distance of $47.35\pm0.15$\,pc
(\citealp{chiappetti18}).  Recent observations
have furthermore shown that HD3167\,c is in a polar orbit which makes
it very likely that also HD3167\,b is in a polar orbit. Thus, this
planet is not only radiatively and tidally heated but also induction
heating is possible. We have obtained high-resolution spectra of
HD3167\,b in- and out-of-transit and searched for emission or
absorption lines tracing volcanic activity and a Mercury-like
exosphere.  Although we derived only upper limits, these
are basically similar to fluxes of the NaD and Ca{\small II}\,H\&K lines in
55\,Cnc\,e obtained by \cite{ridden16}. An interesting aspect of the
detection of the Ca{\small II}\,H\&K line in 55\,Cnc\,e is that it was
only detected in one of the five transits observed. It could thus well
be that the same lines could also be detected in HD3167\,b if more
observations are conducted. We conclude that the most likely
atmosphere of HD3167\,b is a mineral atmosphere produced by the
melting of the rocky surface by intense stellar radiation. Induction
heating can lead to some outgassing if the stellar magnetic
field is strong enough. However, volcanic outgassing is
likely to be quite small because of the relatively large mass of this planet
which leaves the mineral atmosphere produced by the melting of the surface
rocks by intense stellar radiation as the most probable mechanism of the 
atmosphere formation at HD3167\,b. 
Hopefully future observations can help us to found out what
kind of atmosphere of HD3167\,b has.

\section*{Acknowledgements}

We are very thankful to the ESO-staff for carrying out the
observations in service mode, and for providing the community with all
the necessary tools for reducing and analyzing the data.  This work
was generously supported by the Th\"uringer Ministerium f\"ur
Wirtschaft, Wissenschaft und Digitale Gesellschaft. This research has
made use of the SIMBAD database, operated at CDS, Strasbourg, France.
This work has made use of data from the European Space Agency (ESA)
mission {\it Gaia} (\url{https://www.cosmos.esa.int/gaia}), processed
by the {\it Gaia} Data Processing and Analysis Consortium
(DPAC,\url{https://www.cosmos.esa.int/web/gaia/dpac/consortium}). Funding
for the DPAC has been provided by national institutions, in particular
the institutions participating in the {\it Gaia} Multilateral
Agreement. KK acknowledges the support from the Austria Science Fund
(FWF) NFN project S116-N16 and the subproject S11607-N16. The authors
are grateful to Prof. Lena Noack for providing the electrical
conductivity profile of HD\,3167b. We also acknowledge the use of the
APPLAUSE data-base. Funding for APPLAUSE has been provided by DFG
(German Research Foundation, Grant), Leibniz Institute for
Astrophysics Potsdam (AIP), Dr.-Remeis Sternwarte Bamberg (University
N\"urnberg/Erlangen), the Hamburger Sternwarte (University of Hamburg)
and Tartu Observatory. Plate material also has been made available
from Th\"uringer Landessternwarte Tautenburg.












\bsp	
\label{lastpage}
\end{document}